\documentclass[11pt]{article}

\usepackage{epsf,epsfig}
\usepackage{latexsym,amsfonts,amsmath,amssymb,float,rotating}
\usepackage[nolsts]{endfloat}
\usepackage{fullpage,subfigure}
\usepackage{psfrag,fancybox}
\usepackage[dvips]{color}
\usepackage{graphicx}
\usepackage[sort&compress]{natbib}

\newcommand{\comment}[1]{}

\newtheorem{lemma}{Lemma}[section]             
\newtheorem{corollary}{Corollary}[section]     

\title{\bf Inference for stochastic volatility models using time change transformations}
\author{\normalfont Konstantinos Kalogeropoulos \\
\normalfont\it University of Cambridge,
Department of Engineering - Signal Processing Lab
\and \normalfont Gareth O. Roberts \\
\normalfont\it University of Warwick, Department of Statistics
\and \normalfont Petros Dellaportas\\
\normalfont\it Athens University of Economics and Business,
Department of Statistics}

\begin{document}
\maketitle

\begin{abstract}
We address the problem of parameter estimation for diffusion
driven stochastic volatility models through Markov chain Monte
Carlo (MCMC). To avoid degeneracy issues we introduce an
innovative reparametrisation defined through transformations that
operate on the time scale of the diffusion. A novel MCMC scheme
which overcomes the inherent difficulties of time change
transformations is also presented. The algorithm is fast to
implement and applies to models with stochastic volatility. The
methodology is tested through simulation based experiments and
illustrated on data consisting of US treasury bill rates.

\end{abstract}

\section{Introduction}
\label{intro}

Diffusion processes provide natural models for continuous time
phenomena. They are used extensively in diverse areas such as
finance, biology and physics. A diffusion process is defined
through a stochastic differential equation (SDE)

\begin{equation}
\label{eq:standard SDE}
dX_t=\mu(t,X_t,\theta)dt+\sigma(t,X_t,\theta)dW_t,\;0\leq t \leq
T,
\end{equation}

\noindent where W is standard Brownian motion. The drift $\mu(.)$
and volatility $\sigma(.)$ reflect the instantaneous mean and
standard deviation respectively. In this paper we assume the
existence of a unique weak solution to (\ref{eq:standard SDE}),
which translates into some regularity conditions (locally
Lipschitz with a linear growth bound) on $\mu(.)$ and $\sigma(.)$;
see chapter 5 of \cite{rog:wil94} for more details.

The task of inference for diffusion processes is particularly
challenging and has received remarkable attention in the recent
literature; see \cite{sor04} for an extensive review. The main
difficulty is inherent in the nature of diffusions which are
infinite dimensional objects. However, only a finite number of
points may be observed and the marginal likelihood of these
observations is generally unavailable in closed form. This has
stimulated the development of various non-likelihood approaches
which use indirect inference \citep{gou:mon:ren93}, estimating
functions \citep{bib:sor95}, or the efficient method of moments
\citep{gal:tau96}; see also \cite{gal:lon97}.

Most likelihood based methods approach the likelihood function
through the transition density of (\ref{eq:standard SDE}). Denote
the observations by $Y_k$, $k=0,\dots,n$, and with $t_k$ their
corresponding times. If the dimension of $Y_k$ equals that of $X$
(for each k) we can use the Markov property to write the
likelihood, given the initial point $Y_0$, as:

\begin{equation}
\label{eq:transdenisty}
\mathcal{L}(Y,\theta|Y_0)=\prod_{k=1}^{n}p_k(Y_{k}|Y_{k-1};\theta,\Delta),\;\;\Delta=t_{k}-t_{k-1}
\end{equation}

\noindent The transition densities $p_k(.)$ are not available in
closed form but several approximations are available. They may be
analytical, see \cite{ait02}, \cite{ait05}, or simulation based,
see \cite{ped95}, \cite{dur:gal02}. They usually approximate the
likelihood in a way so that the discretisation error can become
arbitrarily small, although the methodology developed in
\cite{bes:pap:rob:f06} succeeds exact inference in the sense that
it allows only for Monte Carlo error. A potential downside of
these methods may be their dependence on the Markov property. In
many interesting multidimensional diffusion models the observation
regime is different and some of their components are not observed
at all.

A famous such example is provided by stochastic volatility models,
used extensively to model financial time series such as equity
prices \citep{hul:whi87,ste:ste91,hes93}, or interest rates
\citep{and:lun98,gal:tau98,dur02}. A stochastic volatility model
is usually represented by a 2-dimensional diffusion

\begin{equation}
\label{eq:sv models}
\left(\begin{array}{ccc}dX_{t}\\
d\alpha_{t}\end{array}\right)=\left(\begin{array}{ccc}\mu_{x}(X_t,\alpha_{t},\theta)\\
\mu_{\alpha}(\alpha_{t},\theta)\end{array}\right)dt+\left(\begin{array}{ccc}\sigma_{x}(\alpha_{t},\theta)
& 0\\ 0 &
\sigma_{\alpha}(\alpha_{t},\theta)\end{array}\right)\left(\begin{array}{ccc}dB_{t}\\
dW_{t}\end{array}\right),
\end{equation}

\noindent where $X$ denotes the observed equity (stock) log-price
or the short term interest rate with volatility $\sigma_{x}(.)$,
which is a function of a latent diffusion $\alpha$. For the
diffusion in (\ref{eq:sv models}), the Markov property may no
longer hold; the distribution of a future stock price depends
(besides the current price) on the current volatility which in
turn depends on the entire price history. Stochastic volatility
models are used

An alternative approach to the problem adopts Bayesian inference
utilizing Markov chain Monte Carlo (MCMC) methods. Adhering to the
Bayesian framework, a prior $p(\theta)$ is first assigned on the
parameter vector $\theta$. Then, given the observations $Y$,
the posterior $p(\theta|Y)$ can be explored through data augmentation
\citep{tan:won87}, treating the unobserved paths of $X$ (paths between
observations) as missing data. The resulting algorithm alternates
between updating $\theta $ and $X$. Initial MCMC schemes following
this programme
were introduced by \cite{jon99}; see also \cite{jon03},
\cite{era01} and \cite{ele:ch:she01}. However, as noted in the
simulation based experiment of \cite{ele:ch:she01} and established
theoretically by \cite{rob:str01}, any such algorithm's convergence
properties will
degenerate as the number of imputed points increases. The problem
may be overcome with the reparametrisation of \cite{rob:str01}, and
this scheme may be applied in all one-dimensional and some
multi-dimensional contexts.  However
this framework does not cover general multidimensional diffusion
models. \cite{ch:pit:she05} and \cite{kal07} offer appropriate
reparametrisations but only for a class of stochastic volatility
models. Alternative reparametrisations were introduced in
\cite{gol:wil07};
see also \cite{gol:wil06} for a sequential
approach.

In this paper we introduce a novel reparametrisation that, unlike
previous MCMC approaches, operates on the time scale of the
observed diffusion rather than its path. This facilitates the
construction of irreducible and efficient MCMC schemes, designed
appropriately to accommodate the time change of the diffusion
path. Our approach is general enough to cover almost every
stochastic volatility model used in practice. The paper is
organized as follows: Section \ref{sec:need for trans} elaborates
on the need for a transformation of the diffusion to avoid
problematic MCMC algorithms. In Section \ref{sec:time change} we
introduce time change transformations whereas Section
\ref{sec:mcmc} provides the details for the corresponding
non-trivial MCMC implementation. The proposed methodology of the
paper is tested and illustrated through numerical experiments in
section \ref{sec:simulations}, and on US treasury bill rates in
section \ref{sec:application}. Finally, section
\ref{sec:discussion} concludes and provides some relevant
discussion.

\section{The necessity of reparametrisation}
\label{sec:need for trans}

A Bayesian data augmentation scheme bypasses a problematic
sampling from the posterior $\pi(\theta|Y)$ by introducing a
latent variable $\mathcal{X}$ that simplifies the likelihood
$\mathcal{L}(Y;\mathcal{X},\theta)$. It usually involves the
following two steps:

\begin{enumerate}
\item \tt{Simulate $\mathcal{X}$ conditional on $Y$ and $\theta$.}

\item \tt{Simulate $\theta$ from the augmented conditional
posterior which is \\proportional to
$\mathcal{L}(Y;\mathcal{X},\theta)\pi(\theta)$.}

\end{enumerate}

It is not hard to adapt our problem to this setting. Y represents
the observations of the price process $X$. The latent variables
$\mathcal{X}$ introduced to simplify the likelihood evaluations
are discrete skeletons of diffusion paths between observations or
entirely unobserved diffusions. In other words, $\mathcal{X}$ is a
fine partition of multidimensional diffusion with drift
$\mu_{X}(t,X_t,\theta)$ and diffusion matrix
$$
\Sigma_{X}(t,X_t,\theta)\;=\;\sigma(t,X_t,\theta)\;\times\;\sigma(t,X_t,\theta)^{\prime},
$$
\noindent and the augmented dataset is
$\mathcal{X}_{i\delta},\;i=0,\dots,T/\delta$, where $\delta$
specifies the amount of augmentation. The likelihood can be
approximated via the Euler scheme
$$
\mathcal{L}^{E}(Y;\mathcal{X},\theta)=\prod_{i=1}^{T/\delta}
p(\mathcal{X}_{i\delta}|\mathcal{X}_{(i-1)\delta}),\;
\mathcal{X}_{i\delta}|\mathcal{X}_{(i-1)\delta}\sim
\mathcal{N}\left(\mathcal{X}_{(i-1)\delta}+\delta\mu_{\mathcal{X}}(.),
\delta\Sigma_{\mathcal{X}}(.)\right),
$$
which is known to converge to the true likelihood
$\mathcal{L}(Y;\mathcal{X},\theta)$ for small $\delta$
\citep{ped95}.

Another property of diffusion processes relates
$\Sigma_{\mathcal{X}}(.)$ to the quadratic variation process.
Specifically we know that

$$
\lim_{\delta\rightarrow
0}\sum_{i=1}^{T/\delta}\left(\mathcal{X}_{i\delta}-\mathcal{X}_{(i-1)\delta}\right)
\left(\mathcal{X}_{i\delta}-\mathcal{X}_{(i-1)\delta}\right)^{T}=
\int_{0}^{T}\Sigma_{\mathcal{X}}(s,\mathcal{X}_s,\theta) ds \;\;
a.s.
$$

The solution of the equation above determines the diffusion matrix
parameters. Hence, there exists perfect correlation between these
parameters and $\mathcal{X}$ as $\delta\rightarrow 0$. This has
disastrous implications for the mixing and convergence of the MCMC
chain as it translates into reducibility for $\delta\rightarrow
0$. This issue was first noted by \cite{rob:str01} for scalar
diffusions and also confirmed by the simulation experiment of
\cite{ele:ch:she01}. Nevertheless, it is not an MCMC specific
problem. It turns out that the convergence of its deterministic
analogue, EM algorithm, is problematic when the amount of
information in the augmented data $\mathcal{X}$ strongly exceeds
that of the observations. In our case $\mathcal{X}$ contains an
infinite amount of information for $\delta\rightarrow 0$.

The problem may be resolved if we apply a transformation so that
the algorithm based on the transformed diffusion is no longer
reducible as $\delta\rightarrow 0$. \cite{rob:str01} provide
appropriate diffusion transformations for scalar diffusions. In a
multivariate context this requires a transformation to a diffusion
with unit volatility matrix; see for instance
\cite{kal:del:rob07}. \cite{ait05} terms such diffusions as
reducible and proves the non-reducibility of stochastic volatility
models that obey (\ref{eq:sv models}). The transformations
introduced in this paper follow a slightly different route and
target the time scale of the diffusion. One of the appealing
features of such a reparametrisation is the generalisation to
stochastic volatility models.

\section{Time change transformations}
\label{sec:time change}

For ease of illustration we first provide the time change
transformation and the relevant likelihood function for scalar
diffusion models with constant volatility. Nevertheless, one of
the main advantages of this technique is the applicability to
general stochastic volatility models.

\subsection{Scalar diffusions}
\label{ssec:time_change_scalar}

Consider a diffusion $X$ defined through the following SDE:

\begin{equation}
\label{eq:simpleSDE} dX_{t}=\mu(t,X_{t},\theta)dt+\sigma
dW_t^{X},\;\;0<t<1\;\;\sigma>0.
\end{equation}

\noindent Without loss of generality, we assume a pair of
observations $X_{0}=y_0$ and $X_{1}=y_1$. For more data, note that
the same operations are possible for every pair of successive
observations that are linked together through the Markov property.
We introduce the latent `missing' path of $X$ for $0\leq t\leq 1$,
denoted by $X^{mis}$, so that $X=(y_0,X^{mis},y_1)$. In the spirit
of \cite{rob:str01}, the goal is to write the likelihood for
$\theta$, $\sigma$ with respect to a parameter-free dominating
measure. Using Girsanov's theorem we can get the Radon-Nikodym
derivative between the law of the diffusion X, denoted by
$\mathbb{P}^{X}$, and that of the driftless diffusion $M=\sigma
dW_{t}^{X}$ which represents Wiener measure and is denoted by
$\mathbb{W}^{X}$. We can write

$$
\frac{d\mathbb{P}(X)}{d\mathbb{W}^{X}}=G(t,X,\theta,\sigma)=\exp\left\{
\int_0^T\frac{\mu(s,X_{s},\theta)}{\sigma^2}dX_{s}-
\frac{1}{2}\int_{0}^{T}\frac{\mu(s,X_{s},\theta)^{2}}{\sigma^{2}}ds\right\}.
$$

\noindent By factorizing $\mathbb{W}^{X}=\mathbb{W}_{y}^{X}\times
Leb(y_1)\times f(y_1;\sigma^2)$, where $\;y_1\sim
\mathcal{N}(y_0,\sigma^2)$ and $Leb(.)$ denotes Lebesgue measure,
we obtain

$$
\frac{d\mathbb{P}(X^{mis},y_0,y_1)}{d\left\{\mathbb{W}_{y}^{X}\times
Leb(y)\right\}}=G(t,X,\theta,\sigma)f(y_1;\sigma),
$$

\noindent where clearly the dominating measure depends on
$\sigma$, since it reflects a Brownian bridge with volatility
$\sigma$.

Now consider the time change transformation which first introduces
a new time scale $\eta(t, sigma))$

\begin{equation}
\label{eq:time_change1}
\eta(t,\sigma)=\int_{0}^{t}\sigma^{2}ds=\sigma^{2}t,
\end{equation}

\noindent and then defines the new transformed diffusion $U$ as

$$
U_{t}=\left\{\begin{array}{lll}X_{\eta^{-1}(t,\sigma)},&0\leq t\leq\sigma^{2},\\
M_{\eta^{-1}(t,\sigma)},&t>\sigma^{2}.\end{array}\right.
$$

\noindent The definition for $t>\sigma^2$ is needed to ensure that
$tU$ is well defined for different values of $\sigma^2>0$ which is
essential in the context of a MCMC algorithm. Using standard time
change properties, see for example \cite{oks00}, the SDE for $U$
is

$$
dU_{t}=\left\{\begin{array}{lll}\frac{1}{\sigma^{2}}\mu(t,U_{t},\theta)dt+dW_t^{U}&0\leq t\leq\sigma^{2},\\
dW_t^{U},&t>\sigma^{2},\end{array}\right.
$$

\noindent where $W^{U}$ is another Brownian motion at the time
scale of $U$. By using Girsanov's theorem again, the law of $U$,
denoted by $\mathbb{P}$, is given through its Radon-Nikodym
derivative with respect to the law $\mathbb{W}^U$ of the Brownian
motion $W^{U}$ at the $U-$time scale:

\begin{eqnarray}
\label{eq:Ugirsanov}
\frac{d\mathbb{P}}{d\mathbb{W}^{U}}\;=\;G(t,U,\theta,\sigma)&=
&\exp\left\{\int_{0}^{+\infty}\frac{\mu(s,U_{s},\theta)}{\sigma^2}dU_{s}-
\frac{1}{2}\int_{0}^{+\infty}\frac{\mu(s,U_{s},\theta)^{2}}{\sigma^4}ds\right\} \nonumber\\
&=&\exp\left\{\int_{0}^{\sigma^2}\frac{\mu(s,U_{s},\theta)}{\sigma^2}dU_{s}-
\frac{1}{2}\int_{0}^{\sigma^2}\frac{\mu(s,U_{s},\theta)^{2}}{\sigma^4}ds\right\}.
\end{eqnarray}

\noindent If we condition the Wiener measure on $y$ at the new
time scale, the likelihood can be written with respect to a
Brownian bridge measure $\mathbb{W}_{y}^{U}$ as

$$
d\mathbb{P}(U,y_0,y_1)=G(t,U,\theta,\sigma)f(y_1;\sigma)
d\left\{\mathbb{W}_{y}^{U}\times Leb(y)\right\}.
$$

\noindent However, this Brownian bridge is conditioned on the
event $U_{\sigma^2}=y_1$ and therefore contains the parameter
$\sigma$. For this reason we introduce a second transformation
which applies to both the diffusion's time scale and its path. Define

\begin{eqnarray}
\label{eq:time_change2}
U_{t}^{0}=(\sigma^2-t)Z_{t/\{\sigma^{2}(\sigma^{2}-t)\}},\;\;0\leq
t<
\sigma^2,\\
U_{t}^{0}=U_t-(1-\frac{t}{\sigma^2})y_0-\frac{t}{\sigma^2}y_1,\;\;0\leq
t< \sigma^2. \nonumber
\end{eqnarray}

\noindent Note that this transformation is 1-1. Its inverse is
given by

$$
Z_{t}=\frac{1+\sigma^2 t}{\sigma^2}U_{\sigma^4 t/(1+\sigma^2
t)}^{0},\; 0\leq t <+\infty.
$$

\noindent Applying Ito's formula and using time change properties
we can also obtain the SDE of $Z$ based on another driving
Brownian motion $W^{Z}$ operating at the $Z-$time:

\begin{equation}
\label{eq:ZSDE}
dZ_t=\frac{\mu\left(t,\frac{\sigma^2}{1+t\sigma^2}\nu(Z_t,\sigma),\theta\right)+\nu(Z_t)
\sigma^2}{1+t\sigma^2}\;dt +dW^{Z}_{t},\;0\leq t<+\infty,
\end{equation}

\noindent where $\nu(Z_t,\sigma)=U_t$. This operation essentially
transforms to a diffusion that runs from 0 to $+\infty$ preserving
the unit volatility. We can re-attempt to write the likelihood
using Girsanov theorem and condition the dominating measure on
$y_1$ to obtain $\mathbb{W}_{y}^{Z}$,

\begin{equation}
\label{eq:timchange_like2}
\frac{d\mathbb{P}(Z,y_0,y_1)}{d\left\{\mathbb{W}_{y}^{Z}\times
Leb(y)\right\}}=G(t,Z,\theta,\sigma)f(y_1;\sigma).
\end{equation}

Despite the fact that $G(Z,\theta,\sigma)$ contains integrals
defined in $(0,\;+\infty)$, it is always finite being an $1-1$
transformation of the Radon-Nikodym derivative between
$\mathbb{P}(U)$ and $\mathbb{W}^{U}$ given by
(\ref{eq:Ugirsanov}). Using the following lemma, we prove that
$\mathbb{W}_{y}^{Z}$ is the law of the standard Brownian motion
and hence the likelihood is written with respect to a dominating
measure that does not depend on any parameters.

\lemma \label{lemma:bbridge}Let W be a standard Brownian motion in
$[0,\;+\infty)$. Consider the process defined for $0\leq t \leq T$
$$
B_t=(T-t)W_{t/\{T(T-t)\}}+
(1-\frac{t}{T})y_0+\frac{t}{T}y_1,\;0\leq t< T
$$

\noindent Then B is a Brownian bridge from $y_0$ at time 0 to
$y_1$ at time T.
\bigskip

\normalfont \noindent {\bf Proof:} See \cite[IV.40.1]{rog:wil94}
for the case $y_0=0$, $T=1$. The extension for general $y_0$ and
$T$ is trivial.

\corollary The measure $\mathbb{W}_{y}^{Z}$ is standard Wiener
measure.

\bigskip

\normalfont \noindent {\bf Proof:} Note that $\mathbb{W}_{y}^{U}$
reflects a Brownian bridge from $y_0$ at time 0 to $y_1$ at time T
and we obtained $\mathbb{W}_{y}^{Z}$ by using the transformation
of Lemma \ref{lemma:bbridge}. Since this transformation is $1-1$,
U is a Brownian bridge (under the dominating measure) if and only
if Z is standard Brownian motion.

\noindent Note that $\mathbb{W}_{y}^{Z}$ is the probability law of
the driftless version of the conditional diffusion Z, whereas the
SDE in (\ref{eq:ZSDE}) corresponds to the unconditional version of
$Z$ itself. The conditional SDE of $Z$ is generally not available
but this does not create a problem. For the path updates we may
use the fact that

\begin{equation}
\label{eq:Zupdates}
\frac{d\mathbb{P}}{d\mathbb{W}_{y}^{Z}}(Z|y_0,y_1)=
G(t,Z,\theta,\sigma)\frac{f(y_1;\sigma)}{f^{P}(y_1;\sigma)}\propto
G(t,Z,\theta,\sigma),
\end{equation}

\noindent where $\mathbb{P}_{y}$ is the law of the conditional
version of $Z$ and $f^{P}(.)$ is the density of $y_1$ under
$\mathbb{P}$. Both $\mathbb{P}_{y}$ and $f^{P}(.)$ are generally
unknown but $G(.)$ and $f(.)$, which appear in
(\ref{eq:timchange_like2}) and (\ref{eq:Zupdates}), are available.

\subsection{Stochastic volatility models}
\label{ssec:trans}

Consider the general class of stochastic volatility models with
SDE given by (\ref{eq:sv models}). Without loss of generality, we
may assume a pair of observations ($X_0=y_0$, $X_{1}=y_1$) due to
the Markov property of the 2-dimensional diffusion $(X,\alpha)$.
The likelihood can then be divided into two parts: The first
contains the marginal likelihood of the diffusion $\alpha$ and the
remaining part corresponds to the diffusion $X$ conditioned on the
path of $\alpha$

$$
\mathbb{P}_{\theta}(X,\alpha)=\mathbb{P}_{\theta}(\alpha)\mathbb{P}_{\theta}(X|\alpha).
$$

\noindent Denote the marginal likelihood for $\alpha$ by
$\mathcal{L}_{\alpha}(\alpha,\theta)$. To overcome reducibility
issues arising from the paths of $\alpha$ one may use the
reparametrisations of \cite{ch:pit:she05} or \cite{kal07}. The
relevant transformations of the latter are

$$
\beta_t= h(\alpha_{t},\theta),\;\frac{\partial
h(\alpha_{t},\theta)}{\partial \alpha_t}\;
=\;\left\{\sigma_{\alpha}(\alpha_{t},\theta)\right\}^{-1},
$$
$$
\gamma_t=\beta_t-\beta_0,\;\;\beta_t=\eta(\gamma_t),
$$
\noindent and the marginal likelihood for the transformed latent
diffusion $\gamma$ becomes

\begin{equation}
\label{eq:svl_likelihood1}
\mathcal{L}_{\gamma}(\gamma,\theta)=\frac{d\mathbb{P}}{d\mathbb{P}}(\gamma)=
G\left\{\eta(\gamma),\theta\right\}.
\end{equation}

\noindent By letting $\alpha_t=g_t^{\gamma}=
h^{-1}(\eta(\gamma_t),\theta)$, the SDE of X conditional on
$\gamma$ becomes:

$$
dX_t=\mu_{x}(X_t,g_t^{\gamma},\theta)dt+\sigma_{x}(g_t^{\gamma},\theta)dB_t,\;\;0\leq
t\leq 1.
$$

\noindent Given the paths of the diffusion $\alpha$, the
volatility function $\sigma_{x}(g_t^{\gamma},\theta)$ may be
viewed as a deterministic function of time. The situation is
similar to that of the previous section. We can introduce a new
time scale

$$
\eta(t,\gamma,\theta)=\int_{0}^{t}\sigma_{x}^{2}(g_t^{\gamma},\theta)ds,
$$
$$
T=\eta(t_k,\gamma,\theta),
$$

\noindent and define $U$ with the new time scale as before ($M$ is
a Brownian motion on the $U-$time scale)

\begin{equation}
\label{eq:time_change1_svol}
U_{t}=\left\{\begin{array}{lll}X_{\eta^{-1}(t)},&0\leq t\leq T,\\
M_{\eta^{-1}(t)},&t>T. \end{array}\right.
\end{equation}

\noindent The SDE for $U$ now becomes

$$
dU_{t}=\left\{\frac{\mu_{x}\left(U_t,\gamma_{\eta^{-1}(t,\gamma,\theta)},\theta\right)}
{\sigma_{x}^{2}(\gamma_{\eta^{-1}(t,\gamma,\theta)},\theta)}\right\}dt+dW_{t}^{U},\;\;0\leq
t\leq T.
$$

\noindent We obtain the Radon Nikodym derivative between the
distribution of $U$ with respect to that of the Brownian motion
$W^{U}$,

$$
\frac{d\mathbb{P}}{d\mathbb{W}^U}\;=\;G(U,\gamma,\theta),
$$

\noindent and introduce $\mathbb{W}_{y}^{U}$ as before. The
density of $y_1$ under $\mathbb{W}^{U}$, denoted by
$f(y,\gamma,\theta)$, is just

$$
f(y_1;\gamma,\theta)\equiv N(y_0,T).
$$

The dominating measure $\mathbb{W}_{y}^{U}$ reflects a Brownian
motion conditioned to equal $y$ at a parameter depended time
$T=\eta(t_{k+1},\gamma,\theta)$. To remove this dependency we
introduce a second time change

\begin{eqnarray}
\label{eq:time_change2_svol}
U^{0}_{t}=(T-t)Z_{t/\{T(T-t)\}},\;0\leq t< T,\\
U^{0}_{t}=U_{t}-(1-\frac{t}{T})y_{0}-\frac{t}{T}y_{1},\;0\leq t<
T.\nonumber
\end{eqnarray}

\noindent Therefore, the SDE for $Z$ is now given by

$$
dZ_{t}=\frac{T}{1+tT}\left\{\frac{\mu\left(\frac{T}{1+tT}\nu(Z_t),\gamma_{k(t,\gamma,\theta)},\theta\right)}
{\sigma_{x}^{2}(\gamma_{k(t,\gamma,\theta)},\theta)}+\nu(Z_t)\right\}dt+dW_{t}^{Z},\;\;0\leq
t<\infty,
$$

\noindent where $k(t,\gamma,\theta)$ denotes the initial time
scale of $X$ and $\nu(Z_t)=U_t$.

\noindent Conditional on $\gamma$, the likelihood can be written
in a similar manner as in (\ref{eq:timchange_like2}):

\begin{equation}
\label{eq:svl_likelihood2}
\frac{d\mathbb{P}}{d\left\{\mathbb{W}_{y}^{Z}\times
Leb(y)\right\}}(Z|y_0,y_1,\gamma)=G(Z,\gamma,\theta)f(y_1;\gamma,\theta)
\end{equation}
\bigskip

\noindent It is not hard to see that $\mathbb{W}_{y}^{Z}$ reflects
a standard Wiener measure and therefore the dominating measure is
independent of parameters. To obtain the full likelihood we need
to multiply the two parts given by (\ref{eq:svl_likelihood1}) and
(\ref{eq:svl_likelihood2}).

\subsection{Incorporating leverage effect}
\label{ssec:leverage}

In the previous section we made the assumption that the increments
of $X$ and $\gamma$ are independent, in other words we assumed no
leverage effect. This assumption can be relaxed in the following
way: In the presence of a leverage effect $\rho$, the SDE of $X$
conditional on $\gamma$ can be written as

$$
dX_t=\mu_{x}(X_t,g_t^{\gamma},\theta)dt+\rho\sigma_{x}(g_t^{\gamma},\theta)dW_t
+\sqrt{1-\rho^2}\sigma_{x}(g_t^{\gamma},\theta)dB_t,\;\;0\leq
t\leq t_k,
$$

\noindent where $W$ is the driving Brownian motion of $\gamma$).
Note that given $\gamma$, $W$ can be regarded as a function of
$\gamma$ and its parameters $\theta$. Therefore, the term
$\rho\sigma_{x}(g_t^{\gamma},\theta)dW_t$ can be viewed as a
deterministic function of time, and it can be treated as part of
the drift of $X_t$. However, this operation introduces additional
problems as the assumptions ensuring a weakly unique solution to
the SDE of $X$ are violated. To avoid this issue we introduce the
infinitesimal transformation

$$
X_t=\mathcal{H}(H_t,\rho,\gamma,\theta)=H_t+\int_{0}^{t}\rho\sigma_{x}(g_s^{\gamma},\theta)dW_s,
$$

\noindent which leads us to the following SDE for $H$:

$$
dH_t=\mu_{x}\left\{\mathcal{H}(X_t,\rho,\gamma,\theta),g_t^{\gamma},\theta\right\}dt
+\sqrt{1-\rho^2}\sigma_{x}(g_t^{\gamma},\theta)dB_t,\;\;0\leq
t\leq t_k.
$$

\noindent We can now proceed as before, defining $U$ and $Z$ based
on the SDE of $H$ in a similar manner as in
(\ref{eq:time_change1_svol}) and (\ref{eq:time_change2_svol})
respectively.

\subsection{State dependent volatility}
\label{ssec:extensions}

Consider the family of state dependent stochastic volatility
models where conditional on $\gamma$, the SDE of $X$ may be
written as:

$$
dX_t=\mu_{x}(X_t,g_t^{\gamma},\theta)dt+
\sigma_{1}(g_t^{\gamma},\theta)\sigma_{2}(X_t,\theta)dB_t,\;\;0\leq
t\leq t_k.
$$

\noindent This class contains among others, the models of
\cite{and:lun98}, \cite{gal:tau98}, \cite{dur02}, \cite{era01}. In
order to apply the time change transformations of section
\ref{ssec:trans}, we should first transform $X$ to $\dot{X}_t$,
through $\dot{X}_t=h(X_t,\theta)$, so that it takes the form of
(\ref{eq:sv models}). Such a transformation, which may be viewed
as the first transformation in \cite{rob:str01}, should satisfy
the following differential equation

$$
\frac{\partial h(X_t,\theta)}{\partial
X_t}=\frac{1}{\sigma_{2}(X_t,\theta)}.
$$

The time change transformations for $U$ and $Z$ may then be
defined on the basis of $\ddot{X}$ that will now have volatility
$\sigma_{1}(g_t^{\gamma},\theta)$. The transformation $h(.)$ also
applies to the observations ($\dot{y}_0$, $\dot{y}_1$) which are
now functions of parameters. This would translated in a parameter
dependent likelihood dominating measure, had it not been for the
second step in (\ref{eq:time_change1_svol}) which in this case
acts like the second transformation in \cite{rob:str01}. Note that
the parameters of $\sigma_{2}(X_t,\theta)$ enter the
reparametrised likelihood in two ways: first through the
$f(y;\gamma,\theta)$ which now should include the relevant
Jacobian term, and second through the drift of $Z$ which is
centered at $0$ based on the transformed observations.

\subsection{Multivariate stochastic volatility models}

We may use the techniques of section \ref{ssec:leverage} to define
time change transformations for multidimensional diffusions.
Consider a $d-$dimensional version of the SDE in
(\ref{eq:simpleSDE}) where $\sigma$ now is a $2\times 2$ matrix
($[\sigma]_{ij}=\sigma_{ij}$). As noted in \cite{kal:del:rob07},
the mapping between $\sigma$ and the volatility matrix
$\sigma\sigma^{T}$ should be 1-1 in order to ensure
identifiability of the $\sigma$ parameters. A way to achieve this,
is by allowing $\sigma$ to be the lower triangular matrix that
produces the Cholesky decomposition of $\sigma\sigma^{T}$. For
$d=2$, the SDE of such a diffusion is given by

$$
dX^{\{1\}}_t=\mu(X^{\{1\}}_t,X^{\{2\}}_t,\theta)dt+\sigma_{11}dB_t,
$$
$$
dX^{\{2\}}_t=\mu(X^{\{1\}}_t,X^{\{2\}}_t,\theta)dt+\sigma_{21}dB_t+\sigma_{22}dW_t.
$$

The time change transformations for $X^{\{1\}}$ will be exactly as
in section \ref{ssec:time_change_scalar}. For $X^{\{2\}}$ note
that given $X^{\{1\}}$ the term $\sigma_{21}dB_t$ is now a
deterministic function of time and may be treated as part of the
drift. Thus, we may proceed following the route of the section
\ref{ssec:leverage}.

Similar transformations can be applied for diffusions that have,
or may be transformed to have, volatility functions independent of
their paths. For example we may assume two correlated price
processes with correlation $\rho_x$:
$$
[\sigma]_{11}=\sigma_{x}^{\{1\}}(g_t^{\gamma},\theta),
$$
$$
[\sigma]_{21}=\rho_{x} \sigma_{x}^{\{2\}}(g_t^{\gamma},\theta),
$$
$$
[\sigma]_{22}=\sqrt{1-\rho_x^2}\sigma_{x}^{\{2\}}(g_t^{\gamma},\theta).
$$

\noindent We may proceed in a similar manner for multivariate
stochastic volatility models of general dimension d.

\section{MCMC implementation}
\label{sec:mcmc}

The construction of an appropriate data augmentation algorithm
involves several issues. The time change transformations introduce
three interesting features to the MCMC algorithm which we address
separately: the presence of three time scales; the need to
update diffusion paths that run from 0 to $+\infty$; and the fact that
time scales depend on parameters. For ease of illustration we will
assume the simple
case of a univariate diffusion with constant volatility and a pair
observations ($X_{0}=y_0$ and $X_{1}=y_1$). Extensions and generalisations
of the algorithm for stochastic volatility models are noted where
appropriate.

\subsection{Three time scales}

We introduce $m$ intermediate points of $X$ at equidistant times
between $0$ and $1$, to give
$X=\{X_{i/(m+1)},\;i=0,1,\dots,m+1\}$. In addition, we  make the
assumption that $m$ is large enough for accurate likelihood
approximations and any error induced by the time discretisation is
negligible for the purposes of our analysis.

Given a value of the time scale parameter $\sigma$, we can get the
$U-$time points by applying (\ref{eq:time_change1}) to each one of
the existing points X, so that

$$
U_{\sigma^2i/(m+1)}=X_{i/(m+1)},\; i=0,1,\dots,m+1.
$$

\noindent  Note that it is only the times that change, the values
of the diffusion remain intact. In a stochastic volatility model
we would use the quantities
$$
\int_{\frac{i}{m+1}}^{\frac{i+1}{m+1}}\sigma_{x}^2(.)ds
$$
for each pair of consecutive imputed points.

The points of $Z$ are multiplied by a time factor which corrects
the deviations from unit volatility. The $Z-$time points may be
obtained by

$$
t^{Z}_{i}=\frac{\sigma^2i/(m+1)}{\sigma^2(\sigma^2-\sigma^2i/(m+1))},\;
i=0,1,\dots,m.
$$

\noindent Clearly this does not apply to the last point which
occurs at time $+\infty$. Therefore, the paths of X, or U, are
more convenient and may be used for likelihood evaluations
exploiting the fact that the relevant transformations are 1-1.
However, the component of the relevant Gibbs sampling scheme is
the diffusion $Z$.

\begin{figure}
\includegraphics[width=6in,height=7in]{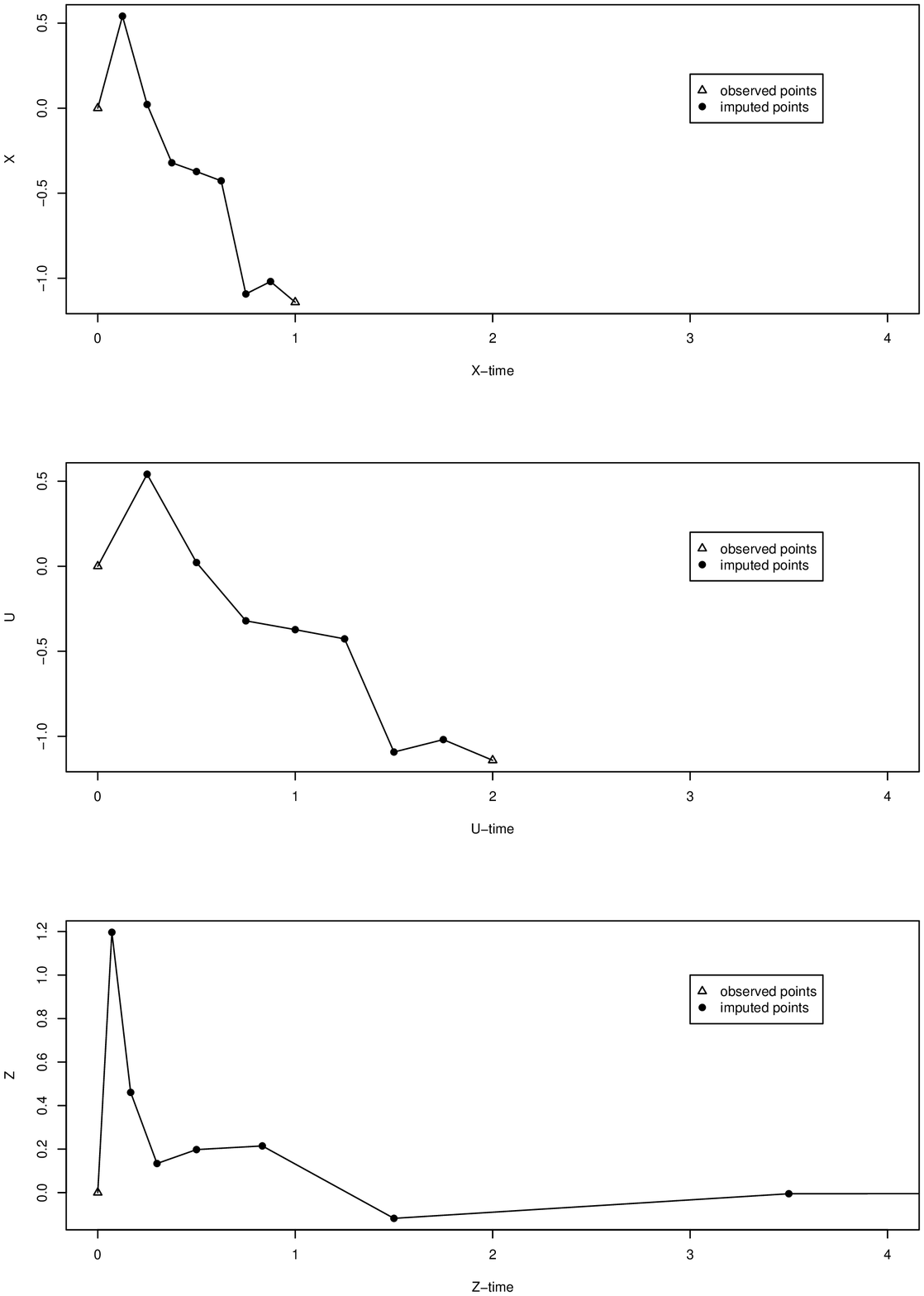}
\caption{Plots of a sample path for $X$, $U$ and $Z$ against their
corresponding times for $\sigma=\sqrt{2}$ and $m=7$. $Z$ equals 0
at time $+\infty$.} \label{fig:timchange1}
\end{figure}

Figure \ref{fig:timchange1} shows a graphical representation of
$X$, $U$ and $Z$ plotted against their corresponding time scales
for $\sigma=\sqrt{2}$ and $m=7$. Although $X$ and $U$ have the
same values, their volatilities are $\sqrt{2}$ and $1$
respectively. The ending point of $Z$ does not appear on the graph
as it occurs at time $+\infty$.

\subsection{Updating the paths of Z}

The paths of Z may be updated using an independence sampler with
the reference measure as a proposal. Here $\mathbb{W}^{Z}$
reflects a Brownian motion at the $Z-$time which is fixed given
the current values of the time-scale parameter(s). An appropriate
algorithm is given by the following steps.

\begin{itemize}
\tt{

\item{Step 1:} Propose a Brownian motion on the $Z-$time, say
$Z^{*}$. The value at the endpoint (time $+\infty$) is not needed.

\item{Step 2:} Transform back to $U^{*}$, using
(\ref{eq:time_change2}).

\item{Step 3:} Accept with probability:
$\min\left\{1,\frac{G(U^*,\theta,\sigma)}
{G(U^*,\theta,\sigma)}\right\}.$

}
\end{itemize}

\subsection{Updating time scale parameters}
\label{ssec:retrospective}

The updates of parameters that define the time scale, such as
$\sigma$, are of particular interest. In almost all cases, their
conditional posterior density is not available in closed form, and
Metropolis steps are inevitable. However, the proposed values of
these parameters will imply different $Z-$ time scales. In other
words, for each potential proposed value for $\sigma$ there exists
a different set of $Z-$ points needed for accurate approximations
of the likelihood the Metropolis accept-reject probabilities. In
theory, this would pose no issues had we been able to store an
infinitely thin partition of $Z$, but of course this is not
possible.

We use retrospective sampling ideas; see \cite{pap:rob05} and
\cite{bes:rob05} for applications in different contexts. Under the
assumption of a sufficiently fine partition of $Z$, all the
non-recorded intermediate points contribute nothing to the
likelihood and they are irrelevant in that respect; the set of
recorded points is sufficient for likelihood approximation
purposes.  Alternatively, we may argue that their distribution is
given by the likelihood reference measure which reflects a
Brownian motion. Thus, they can be drawn after the proposal of the
candidate value of the time scale parameter. To ensure
compatibility with the recorded partition of $Z$, it suffices to
condition on their neighboring points. This is easily done using
standard Brownian bridge properties: Suppose that we want to
simulate the value of $Z$ at time $t_b$ which fall between the
recorded values at times $t_a$ and $t_c$, so that $t_a \leq t_b
\leq t_c$. Denote by $Z_{t_{a}}$ and $Z_{t_{c}}$ the corresponding
$Z$ values. Under the assumption that $Z$ is distributed according
to $\mathbb{W}_{y}^{Z}$ between $t_a$ and $t_c$ we have that

\begin{equation}
\label{eq:rettrosp_bridges} Z_{t_b}|\;Z_{t_a},Z_{t_c}\sim
N\left\{\frac{(t_{b}-t_{a})Z_{t_{c}}+ (t_{c}-t_{b})Z_{t_{a}}}
{t_{c}-t_{a}},\;\;\frac{(t_{b}-t_{a})(t_{c}-t_{b})}{t_{c}-t_{a}}\right\}.
\end{equation}

\begin{figure}
\includegraphics[width=6in,height=7in]{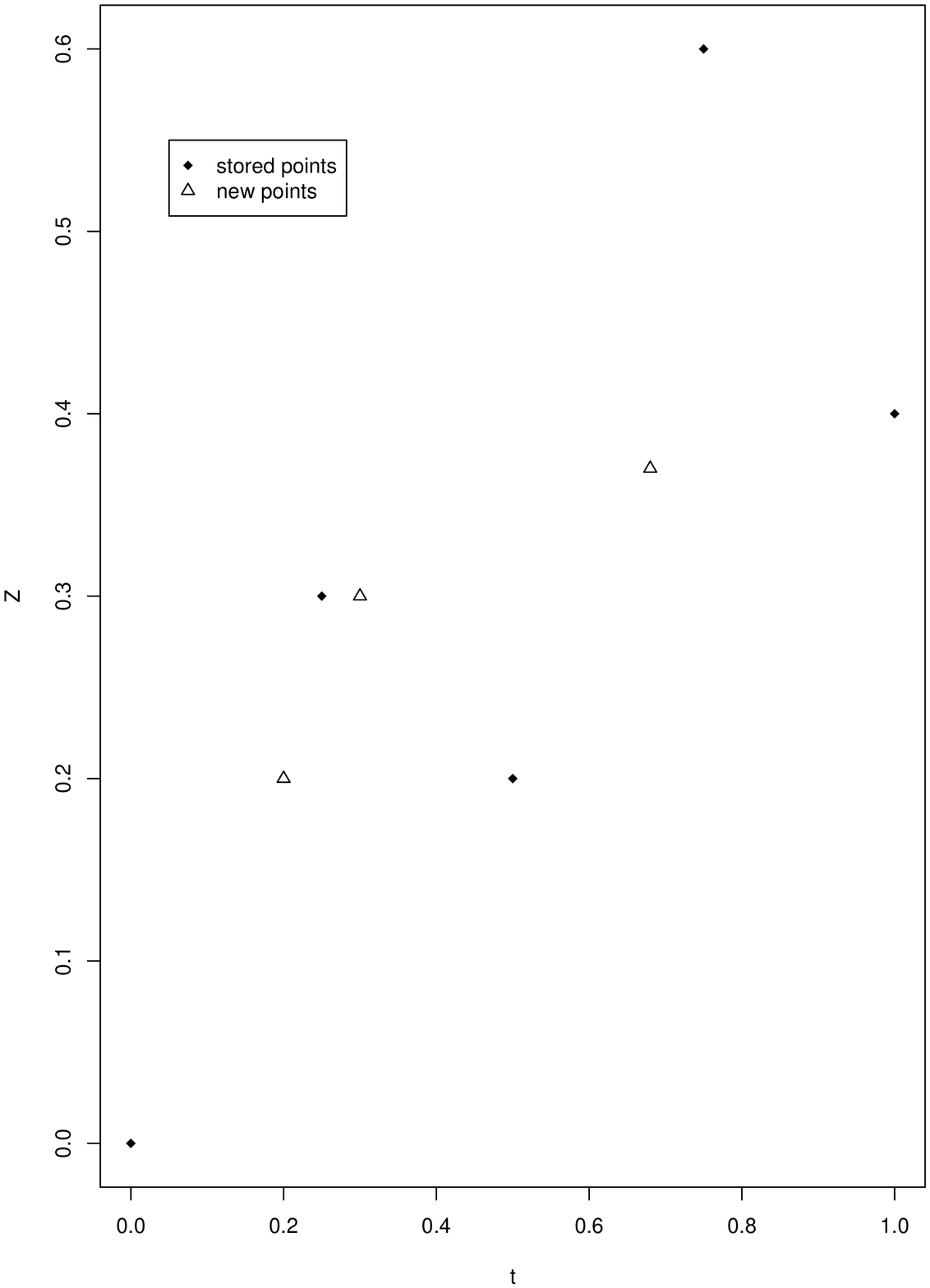}
\caption{Updates of time scale parameters: For every proposed
value of them, new points are required and should obtained
conditional on the stored points.} \label{fig:timchange2}
\end{figure}

\noindent The situation is pictured in Figure
\ref{fig:timchange2}, where the black bullets represent the stored
points and the triangles the new points required for a proposed
value of $\sigma$. The latter should be drawn retrospectively
given the former via (\ref{eq:rettrosp_bridges}).

\noindent A suitable algorithm for the $\sigma-$updates may be
summarized through the following steps:

\begin{itemize}
\tt{

\item{Step 1:} Propose a candidate value for $\sigma$, say
$\sigma^{*}$.

\item{Step 2:} Repeat for each pair of successive points:
\begin{itemize}

\item Use (\ref{eq:time_change1}) and (\ref{eq:time_change2}) to
get the new times associated with it.

\item Draw the values of $Z$ at the new times using
(\ref{eq:rettrosp_bridges}).

\item Transform back to $U^{*}$, using (\ref{eq:time_change2}).
\end{itemize}

Form the entire path $U^{*}$ by appropriately joining its bits.

\item{Step 3:} Accept with probability:
$\min\left\{1,\frac{G(U^*,\theta,\sigma^{*})f(y;\sigma^{*})}
{G(U^*,\theta,\sigma)f(y;\sigma)}\right\}.$
}

\end{itemize}

Note that in a stochastic volatility model the paths of the
transformed diffusion $\gamma_t$ are associated with the time
scale of the $Z$. Therefore a similar algorithm may be used for
their updates.

\section{Simulations}
\label{sec:simulations}

As discussed in section \ref{sec:need for trans}, appropriate
reparametrisations are necessary to avoid issues regarding the
mixing and convergence of the MCMC algorithm. In fact, the chain
becomes reducible as the level of augmentation increases. This is
also verified by the numerical examples performed in \cite{kal07}
even in very simple stochastic volatility models. In this section
we perform a simulation based experiment to check the immunity of
MCMC schemes to increasing levels of augmentation, as well as the
ability of our estimation procedure to retrieve the correct values
of the diffusion parameters despite the fact that the series is
partially observed at only a finite number of points. We simulate
data from the following stochastic volatility model

\begin{eqnarray*}
dX_t&=&\kappa_{x}(\mu_{x} - X_t)dt+\rho\exp(\alpha_t/2)dW_t
+\sqrt{1-\rho^2}\exp(\alpha_t/2)dB_t,\\
d\alpha_t&=&\kappa_{\alpha}(\mu_{\alpha}-\alpha_t)dt+\sigma dW_t,
\end{eqnarray*}
\bigskip

\noindent where B and W are independent Brownian motions, and
$\rho$ reflects the correlation between the increments of X and
$\alpha$, also term as leverage effect. A high frequency Euler
approximating scheme with a step of $0.001$ was used for the
simulation of the diffusion paths. Specifically, $500,001$ points
were drawn and one value of $X$ for every $1000$ was recorded,
thus forming a dataset of $501$ observations of $X$ at $0\leq t
\leq 500$. The parameter values were set to $\rho=-0.5$,
$\sigma=0.4$, $\kappa_x=0.2$, $\mu_x=0.1$, $\kappa_{\alpha}=0.3$
and $\mu_{\alpha}=-0.2$

The transformations required to construct an irreducible data
augmentation scheme are listed below. First we transform $\alpha$
to $\gamma$ through

$$
\gamma_t=\frac{\alpha_t-\alpha_0}{\sigma},\; 0\leq t \leq 500,
$$
$$
\alpha_t=\nu(\gamma_t,\sigma,\alpha_0)=\alpha_0+\sigma\gamma_t.
$$
\bigskip

\noindent Given $\gamma$, and for each pair of consecutive
observation times $t_{k-1}$ and $t_{k}$ ($k=1,2,\dots,500$) on X,
we transform as follows: First, we remove the term introduced from
the leverage effect

$$
H_t=X_t-\int_{t_{k-1}}^{t}\rho\exp\left\{\nu(\gamma_s,\sigma,\alpha_0)/2\right\}dW_s,\;t_{k-1}\leq
t\leq t_{k},
$$

\noindent and consequently we set

$$
\eta(t)=\int_{t_{k-1}}^{t}(1-\rho)^2\exp\left\{\nu(\gamma_s,\sigma,\alpha_0\right\}ds.
$$

\noindent Then, $U$ and $Z$ may be defined again from
\ref{eq:time_change1_svol} and \ref{eq:time_change2_svol}
respectively, but based on $H$ rather on $X$. The elements of the
MCMC scheme are $Z$,$\gamma$, $\alpha_0$ and the parameters
$(\kappa_x,\mu_x,\kappa_\alpha,\mu_\alpha,\rho,\sigma)$.

\begin{figure}

\psfrag{Lag - sigma}{Lag - $\sigma$}

\psfrag{Lag - rho}{Lag - $\rho$}

\includegraphics[width=6in,height=7in]{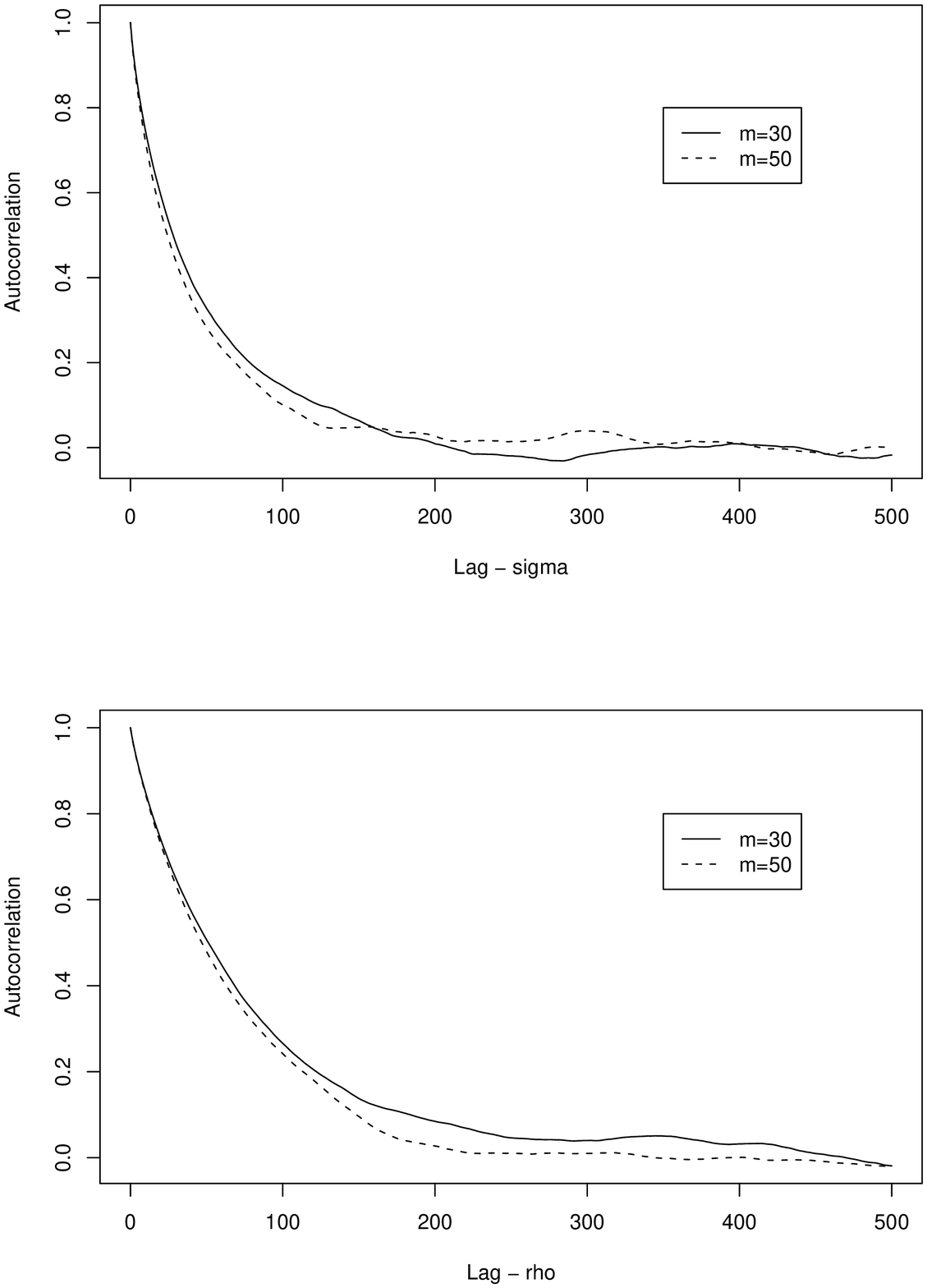}
\caption{Autocorrelation plots for the posterior draws of $\rho$
and $\sigma$ for different numbers of imputed points $m=30,50$.
Simulation example of Chapter 3.} \label{fig:sim1}
\end{figure}

\begin{figure}

\psfrag{kappax}{$\kappa_{x}$}

\psfrag{mux}{$\mu_{x}$}

\psfrag{kappaa}{$\kappa_{\alpha}$}

\psfrag{mua}{$\mu_{\alpha}$}

\psfrag{sigma}{$\sigma$}

\psfrag{rho}{$\rho$}
\includegraphics[width=6in,height=7in]{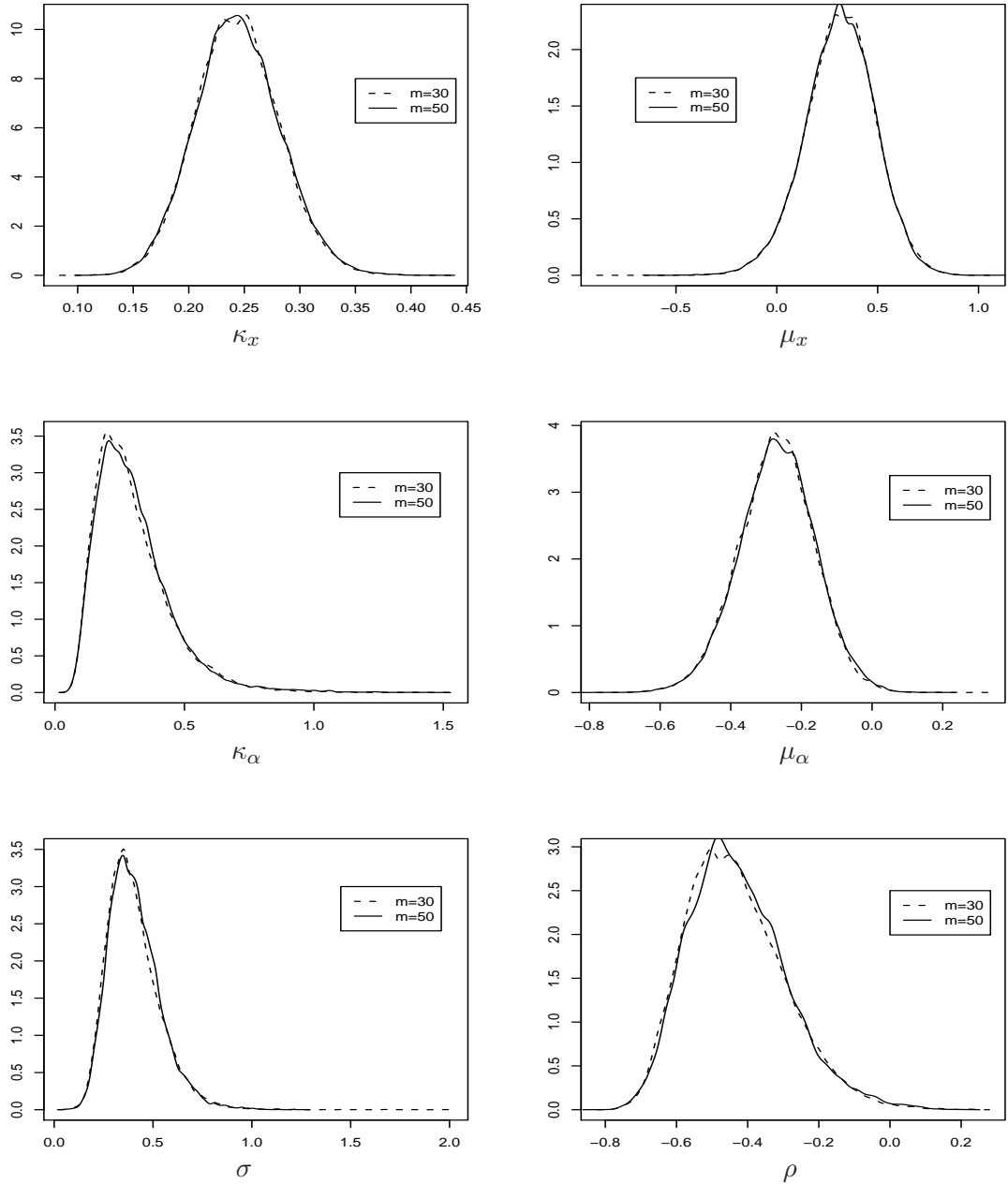}
\caption{Kernel densities of the posterior draws of all the
parameters for different numbers of imputed points $m=30,50$.
Simulation example of Chapter 3.} \label{fig:sim2}
\end{figure}

We proceed by assigning flat priors to all the parameters,
restricting $\kappa_{x}$, $\kappa_{\alpha}$, $\sigma$ to be
positive and $\rho$ to be in $(-1,1)$. The number of imputed
points was set to $30$ and $50$, the length of the overlapping
blocks needed for the updates of $\gamma$ was $2$, and the
relevant acceptance rate $75\%$ whereas the acceptance rate for
$X$ was $95\%$. Figure \ref{fig:sim1} shows autocorrelation plots
for the 2-dimensional diffusion's $(X,\alpha)^{\prime}$ volatility
parameters $\rho$ and $\sigma$. There is no sign of any increase
in the autocorrelation to raise suspicions against the
irreducibility of the chain. Figure \ref{fig:sim2} shows density
plots for all parameters and both values of $m$. These plots
indicate a sufficiently fine discretisation and a good agreement
with true values of the parameters. The latter is also confirmed
by Table \ref{tab:sim1}.

\begin{table}
\begin{tabular}{|c | cccccc|}
\hline  Parameter & True value &  Post. mean  & Post. SD & Post
$2.5\%$& Post median & Post $97.5\%$ \\\hline
$\kappa_{x}$      & 0.2 & 0.244  & 0.038  & 0.173  & 0.243  & 0.321 \\
$\mu_{x}$         & 0.1 & 0.313  & 0.174  & -0.046 & 0.317  & 0.641 \\
$\kappa_{\alpha}$ & 0.3 & 0.304  & 0.148  & 0.110  & 0.277  & 0.672 \\
$\mu_{\alpha}$    & -0.2& -0.268 & 0.107  & -0.484 & -0.267 & -0.059\\
$\sigma$          & 0.4 & 0.406  & 0.130  & 0.202  & 0.390  & 0.705 \\
$\rho$            & -0.5& 0.477  & 0.138  & -0.657 & -0.491 & -0.066\\
\hline
\end{tabular}
\caption{Summaries of the posterior draws for the simulation
example of Chapter 3 for $m=50$.} \label{tab:sim1}
\end{table}

\section{Application: US treasury bill rates}
\label{sec:application}

\begin{figure}
\includegraphics[width=6in,height=7in]{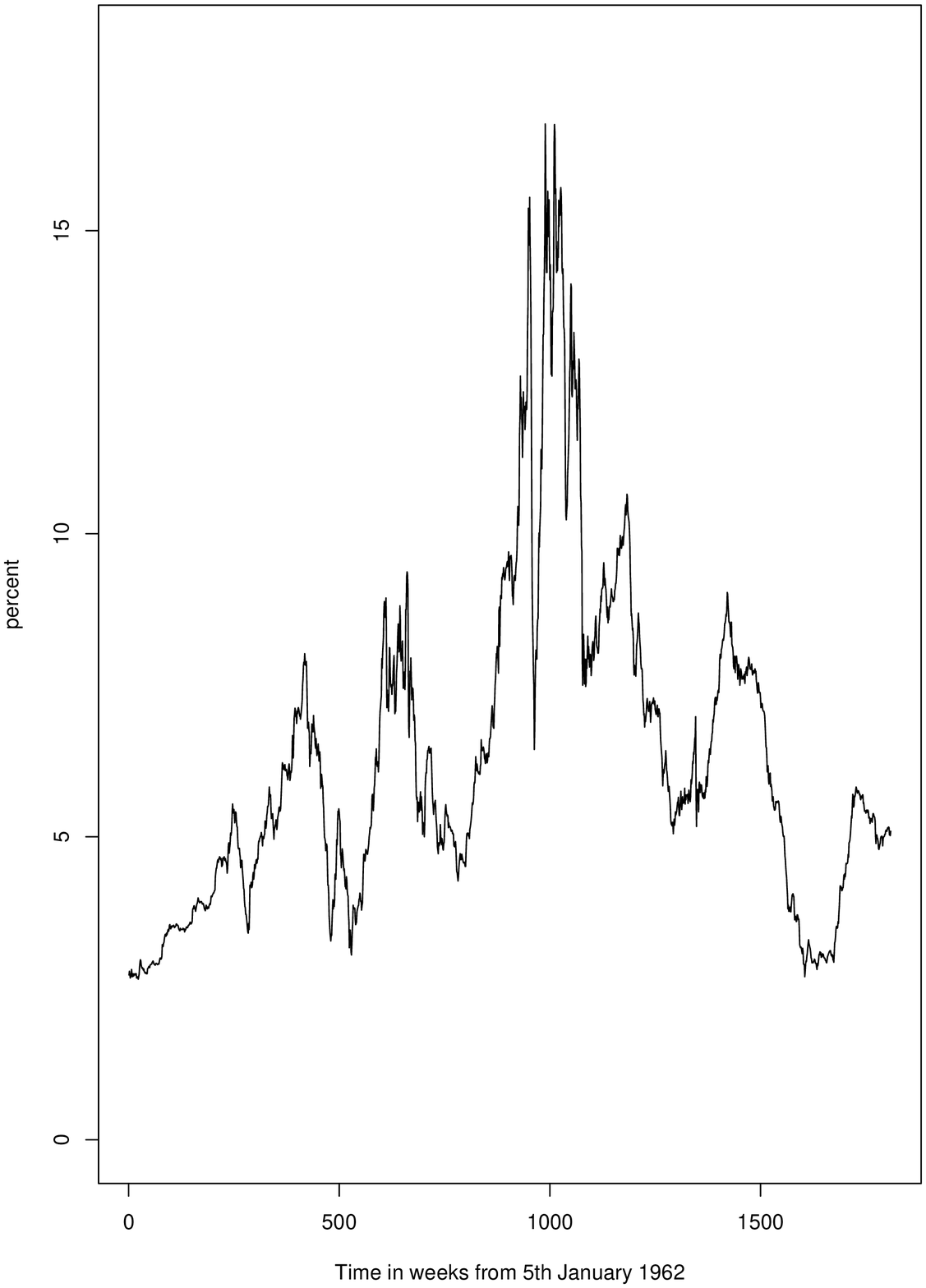}
\caption{Weekly $3-$month US Treasury bill rate from the $5$th of
January $1962$ up to the $30$th of August $1996$.}
\label{fig:exdata}
\end{figure}

To illustrate the time change methodology we fit a stochastic
volatility model to US treasury bill rates. The dataset consists
of $1809$ weekly observations (Wednesday) of the $3-$month US
Treasury bill rate from the $5$th of January $1962$ up to the
$30$th of August $1996$. The data are plotted in Figure
\ref{fig:exdata}.

Previous analyses of these data include \cite{and:lun98},
\cite{gal:tau98}, \cite{dur02}, \cite{dur:gal02}, \cite{era01},
and \cite{gol:wil06}. Apart from some slight deviations the
adopted stochastic volatility models consisted of the following
SDE.

\begin{eqnarray}
\label{eq:exsv1}
dr_t&=&(\theta_{0} - \theta_{1}r_t)dt+r_t^{\psi}\exp(\alpha_t/2)dB_t\nonumber,\\
d\alpha_t&=&\kappa(\mu-\alpha_t)dt+\sigma dW_t,
\end{eqnarray}

\noindent with independent Brownian motions $B$ and $W$. In some
cases the following equivalent model was used:

\begin{eqnarray}
\label{eq:exsv2}
dr_t&=&(\theta_{0} - \theta_{1}r_t)dt+\sigma_{r}r_t^{\psi}\exp(\alpha_t/2)dB_t\nonumber,\\
d\alpha_t&=&-\kappa\alpha_t dt+\sigma dW_t.
\end{eqnarray}

\noindent We proceed with the model in (\ref{eq:exsv1}), as
posterior draws of its parameters exhibit substantially less
autocorrelation. In line with \cite{gal:tau98} and
\cite{gol:wil06}, we also set $\psi=1$. \cite{era01}, \cite{dur02}
and \cite{dur:gal02} assume general `elasticity of variance'
$\psi$ but their estimates do not indicate a significant deviation
from $1$. By setting $X_t=log(r_t)$, the volatility of $X_t$
becomes $\exp(\alpha_t/2)$. Therefore the $U-$time for two
consecutive observation times $t_{k-1}$ and $t_{k}$ is defined as

$$
\eta(t)=\int_{t_{k-1}}^{t}\exp(\alpha_t)ds,
$$

\noindent and $U$ and $Z$ are given by
(\ref{eq:time_change1_svol}) and (\ref{eq:time_change2_svol})
respectively . We also transform $\alpha$ to $\gamma$ as before:

$$
\gamma_t=\frac{\alpha_t-\alpha_0}{\sigma},
$$
$$
\alpha_t=\nu(\gamma_t,\sigma,\alpha_0)=\alpha_0+\sigma\gamma_t.
$$

We applied MCMC algorithms based on $Z$ and $\gamma$ to sample
from the posterior of the parameters $\theta_0$, $\theta_{1}$,
$\kappa$, $\mu$ and $\sigma$. The time was measured in years
setting the distance between successive Wednesdays to $5/252$.
Non-informative priors were assigned to all the parameters,
restricting $\kappa$ and $\sigma$ to be positive to ensure
identifiability and eliminate the possibility of explosion. The
algorithm was run for $50,000$ iterations and for $m$ equal to
$10$ and $20$. To optimize the efficiency of the chain we set the
length of the overlapping blocks of $\gamma$ to $10$ which
produced an acceptance rate of $51.9\%$. The corresponding
acceptance rate for $Z$ was $98.6\%$ .

\begin{figure}

\psfrag{kappa}{$\kappa$}

\psfrag{mu}{$\mu$}

\psfrag{t0}{$\theta_0$}

\psfrag{t1}{$\theta_1$}

\psfrag{sigma}{$\sigma$}

\psfrag{rho}{$\rho$}
\includegraphics[width=6in,height=7in]{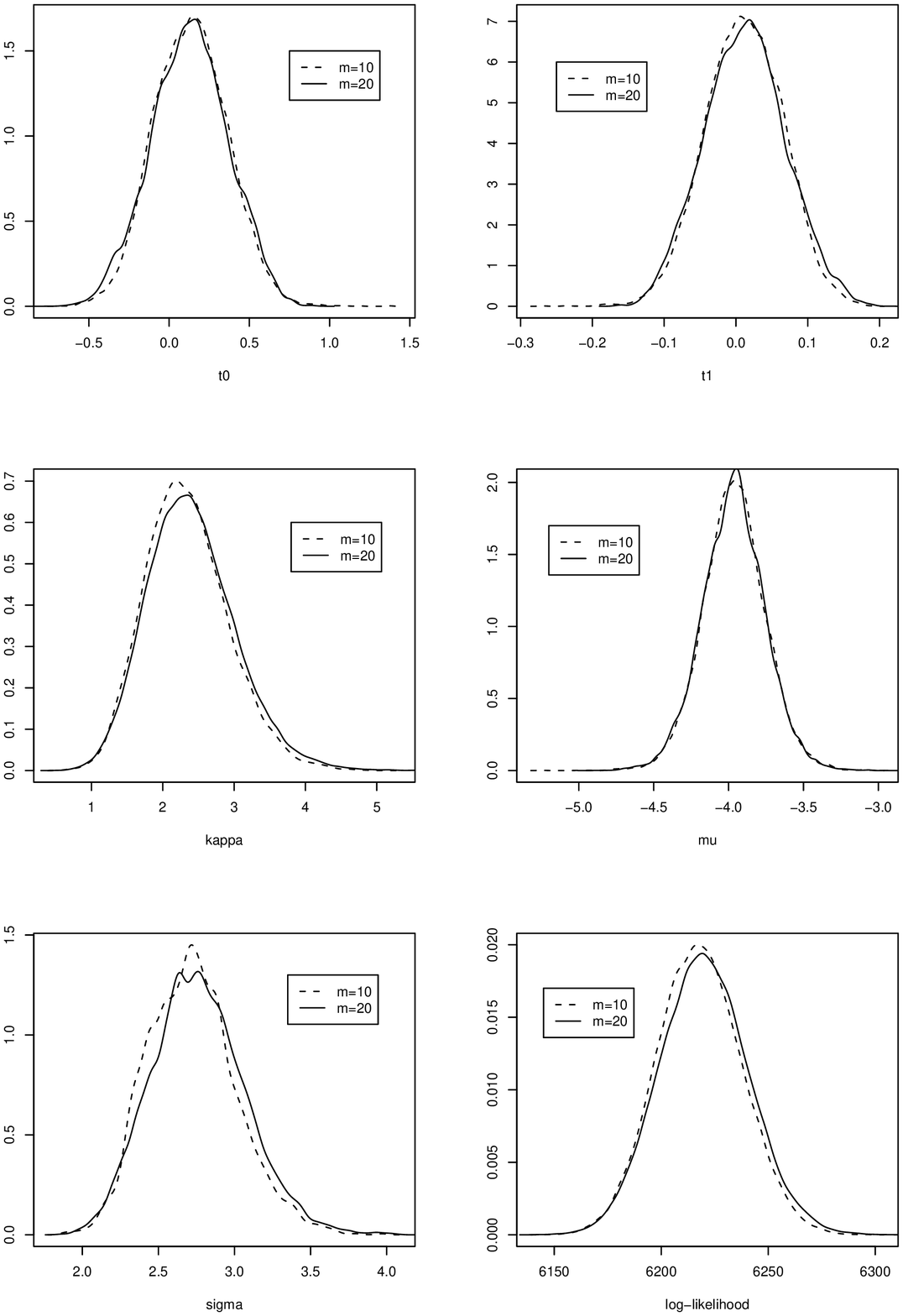}
\caption{Kernel densities of the posterior draws of all the
parameters and the log-likelihood for different values of imputed
points $m=10,20$. Example on Weekly $3-$month US Treasury bill
rates.} \label{fig:ex1}
\end{figure}

\begin{figure}

\psfrag{Lag - sigma}{Lag - $\sigma$}

\psfrag{Lag - kappa}{Lag - $\kappa$}

\psfrag{Lag - mu}{Lag - $\mu$}

\psfrag{Lag - t0}{Lag - $\theta_0$}

\psfrag{Lag - t1}{Lag - $\theta_1$}

\includegraphics[width=6in,height=7in]{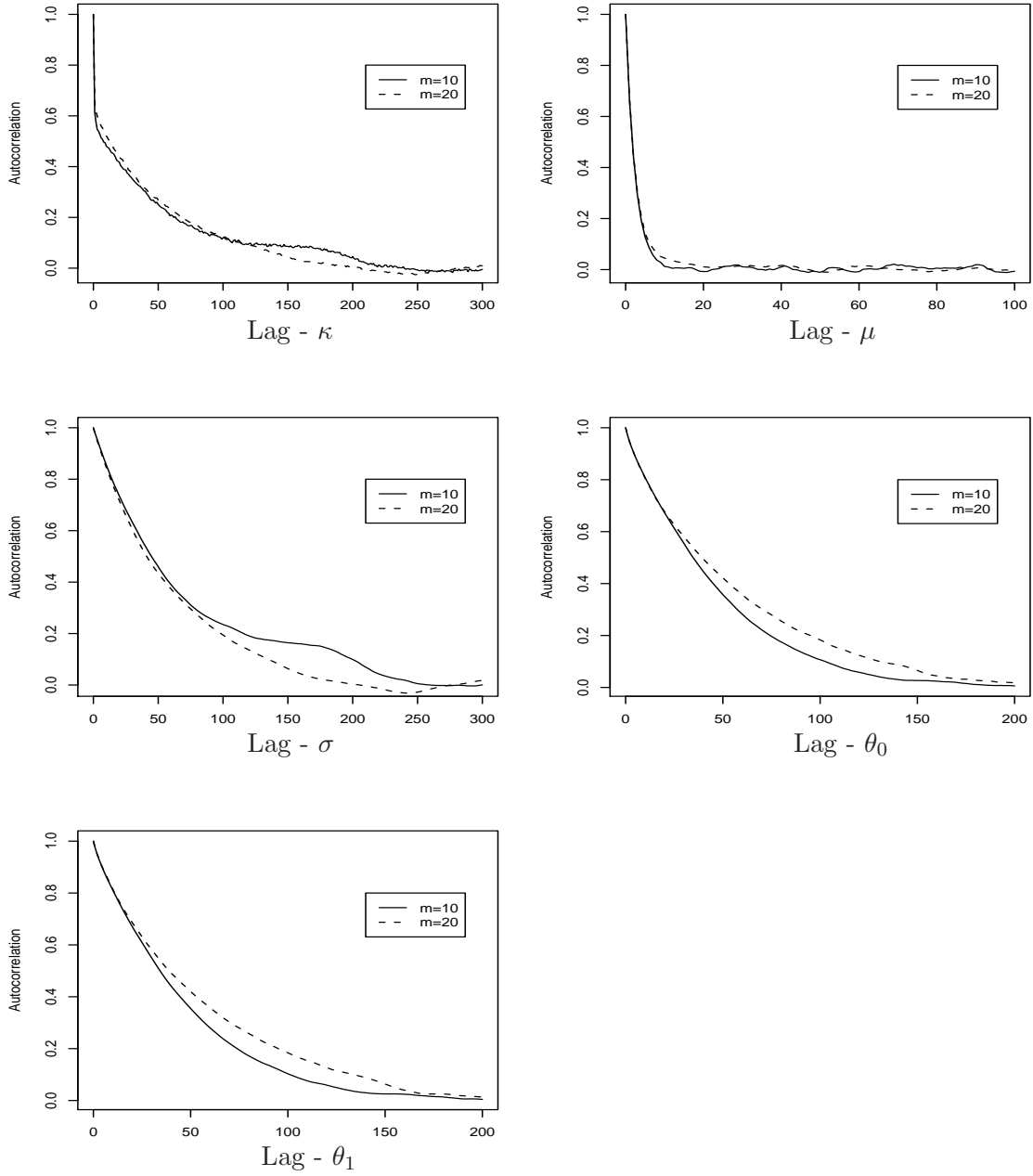}
\caption{Autocorrelation plots for the posterior draws of the
model parameters for different numbers of imputed points $m=10,20$
for the analysis of Weekly $3-$month US Treasury bill rates.}
\label{fig:ex2}
\end{figure}

The kernel density plots of the posterior parameters and
likelihood (Figure \ref{fig:ex1}) indicate that a discretisation
from an $m$ of $10$ or $20$ provide reasonable approximations. The
corresponding autocorrelation plots of Figure \ref{fig:ex2} do not
show increasing autocorrelation in $m$, a feature that would
reveal reducibility issues. Finally, summaries of the posterior
draws for all the parameters are provided in Table \ref{tab:ex1}.
The parameters $\kappa$, $\mu$ and $\sigma$ are different from $0$
verifying the existence of stochastic volatility. On the other
hand, there is no evidence to support the existence of mean
reversion on the rate, as $\theta_0$ and $\theta_{1}$ are not far
from $0$. The results are in line with those of \cite{dur02},
\cite{dur:gal02} and \cite{gol:wil06}.

\begin{table}
\begin{tabular}{|c | ccccc|}
\hline  Parameter &  Post. mean  & Post. SD & Post $2.5\%$& Post
median & Post $97.5\%$ \\\hline
$\theta_{0}$  & 0.130  & 0.238  & -0.347 & 0.132  & 0.589 \\
$\theta_{1}$  & 0.013  & 0.057  & -0.096 & 0.013  & 0.125 \\
$\kappa$      & 2.403  & 0.620  & 1.319  & 2.360  & 3.745 \\
$\mu$         & -3.966 & 0.211  & -4.384 & -3.964 & -3.547\\
$\sigma$      & 2.764  & 0.311  & 2.199  & 2.750  & 3.420 \\
\hline
\end{tabular}
\caption{Summaries of the posterior draws for the stochastic
volatility model of Weekly $3-$month US Treasury bill rates.}
\label{tab:ex1}
\end{table}

\section{Discussion}
\label{sec:discussion}

Data augmentation MCMC schemes constitute a very useful tool for
likelihood-based inference on diffusion models. They may not have
the appealing properties of complete elimination of the time
discretisation error \citep{bes:pap:rob:f06}, or the closed form
approximate likelihood expressions of \cite{ait02}, but
nevertheless they give a satisfactory and very general solution to the
problem.  However data augmentation schemes require careful construction
to avoid the degeneracy issues described at the beginning of this paper.

Here, we introduce an innovative transformation which
operates by altering the time axis of the diffusion.
To accommodate the special features of time change
transformations we also introduce a novel efficient MCMC scheme
which mixes rapidly and is not provibitively computationally
expensive. Our method is also easy to
implement and introduces no additional approximation error other
than that included in methodologies based on a discretisation of
the diffusion path. Moreover it is general enough to include
general stochastic volatility models.

Further work will consider problems with state-dependent
volatility and models which involve jump diffusions, to which the
methodology introduced here can be easily applied. Fundamental to
our approach here has been the introduction of a non-centered
parameterisation to decouple dependence inherent in the model
between missing data and volatility parameters. However
non-centered constructions are not unique, as illustrated by the
choice in the diffusion context between the state rescaling
approaches of \cite{rob:str01,gol:wil07} and the time-stretching
strategy adopted here. Clearly, further work is required to
investigate the relative merits of these approaches in different
situations.

\section{Acknowledgements}
This work was supported by the EPSRC grant GR/S61577/01 and by the
Marie Curie fellowships. The authors would also like to thank
Omiros Papaspiliopoulos for helpful discussions.

\newpage{}
\bibliography{tchange}

\end{document}